\documentclass[prl,twocolumn,showpacs,superscriptaddress,floatfix]{revtex4}
\usepackage{times}
\usepackage{latexsym,amsmath,amssymb,bm,euscript}
\usepackage[dvips]{color}
\usepackage{dsfont}
\usepackage{mathrsfs}
\usepackage{multirow}
\usepackage{graphicx}
\usepackage{amsmath}

\begin{document}

\title{Disorder-induced Floquet Topological Insulators}
\date{\today}
\author{Paraj Titum}
\affiliation{Institute of Quantum Information and Matter, Dept. of Physics, Caltech, Pasadena, CA 91125}
\author{Netanel H. Lindner}
\affiliation{Institute of Quantum Information and Matter, Dept. of Physics, Caltech, Pasadena, CA 91125}
\affiliation{Department of Physics, Technion - Israel Institute of Technology, Haifa 32000, Israel}
\author{Mikael	C. Rechtsman}
\affiliation{Department of Physics, Technion - Israel Institute of Technology, Haifa 32000, Israel}
\author{ Gil Refael }
\affiliation{Institute of Quantum Information and Matter, Dept. of Physics, Caltech, Pasadena, CA 91125}

\begin{abstract}
We investigate the possibility of realizing a disorder-induced topological Floquet spectrum in two-dimensional periodically-driven systems.
Such a state would be a dynamical realization of the topological Anderson insulator. We establish that a disorder-induced trivial-to-topological transition indeed occurs, and characterize it by computing the disorder averaged Bott index, suitably defined for the time-dependent system. The presence of edge states in the topological state is confirmed by exact numerical time-evolution of wavepackets on the edge of the system. We consider the optimal driving regime for experimentally observing  the Floquet topological Anderson insulator, and discuss its possible realization in photonic lattices.
\end{abstract}


\maketitle
Topological states have been an ongoing fascination in condensed matter and recently led to the prediction \cite{Bernevig15122006,PhysRevLett.95.146802,PhysRevLett.98.106803} and realization \cite{Konig02112007,Hsieh2008,Roth17072009} of various topological phases, including topological insulators (TIs).
TIs possess extraordinary properties (gapless edge states \cite{JPSJ.77.031007,PhysRevLett.101.246807}, topological excitations \cite{Qi2008}) and have myriad potential applications from spintronics to topological quantum computation \cite{RevModPhys.80.1083}. One method to generate topological states is via periodic driving of a topologically trivial system out of equilibrium.
These so-called Floquet topological Insulators (FTIs) might be obtained by irradiating ordinary semiconductors with a spin-orbit interaction \cite{Lindner2011,PhysRevB.87.235131},  or graphene-like systems \cite{PhysRevB.84.235108,PhysRevB.79.081406,PhysRevB.88.245422,PhysRevB.89.205408};  analogues in superconducting systems have also been proposed \cite{PhysRevLett.106.220402,PhysRevLett.111.136402}. Topological phases thus obtained introduce new parameters for controlling the phase, such as the frequency and intensity of the drive. Also, while FTIs have gapless edge states (just as topological insulators do), they exhibit phases with no analog in equilibrium systems \cite{PhysRevX.3.031005,PhysRevLett.110.200403}. Remarkably, topological Floquet spectra were recently experimentally realized in artificial photonic lattices where edge transport was observed \cite{Rechtsman2013}, as well as in the solid state \cite{Wang25102013}. The tunability of photonic systems is conducive to exploring a variety of 
effects, including the influence of controlled disorder.

Here, we are interested in the interplay of disorder and topological behavior. In two-dimensional TIs, it has been shown \cite{PhysRevLett.98.076802} that ballistic edge modes are robust to disorder as long as there is a bulk mobility gap. In contrast, disorder completely localizes the states of trivial non-interacting (and spinless) 2-D systems. In the presence of strong spin-orbit coupling, however, disorder can induce a  phase transition from a trivial to a topological Anderson insulator (TAI) phase,  which exhibits quantized conductance at finite disorder strengths.
TAIs were predicted in electronic models \cite{PhysRevLett.102.136806,PhysRevLett.103.196805,PhysRevB.84.035110,PhysRevB.85.195125}, but have not been observed experimentally.

\begin{figure}
\includegraphics[width=\linewidth]{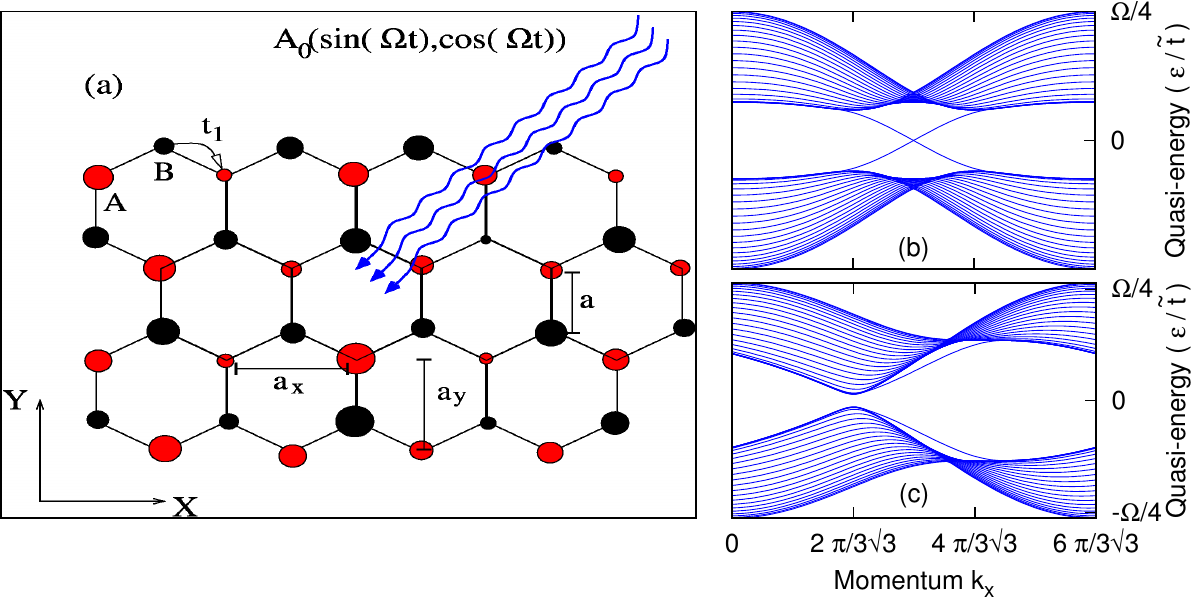}
\caption{(a) Schematic representation of the system indicating uniformly disordered graphene in the presence of a staggered
mass potential and a circularly polarized light. Red-Black coloring indicates the staggered mass in the sublattices A and B, and the variable radius the disorder potential.
(b) The Floquet band structure for the pure system with parameters, $A_0=1.43$,  $M=0$, and
$\Omega/\tilde{t}=12$. The system is topological and supports edge states. The bulk gap is given by the topological mass $\Delta/\tilde{t} \approx 0.75$. (c)
 A trivial Floquet band structure. All parameters are the same as (b) except $M/\tilde{t}=0.85$.
}\label{fig:1}
\end{figure}

Can disorder induce topological phases in trivial periodically-driven systems?
Naively, we would think that disorder would destroy the conditions that give rise to Floquet topological phases.
Nevertheless, we find concrete examples where disorder induces a topological phase.
Here we investigate such transitions in driven systems, and describe their unique properties.
The model we consider is a graphene-like lattice subject to circularly polarized light, with a staggered potential and on-site disorder.   We obtain the phase diagram as a function of disorder strength by calculating the disorder-averaged bulk topological invariant viz., the Bott index. The time-evolution of wavepackets reveals gapless edge modes in the topological phase. As we explain below, our model is especially appealing as it is amenable to experimental realization in photonic lattices.

 Our starting point is the tight binding Hamiltonian of a honeycomb lattice subject to circularly polarized light,
\begin{subequations}
\label{eq:1}
\begin{align}
H_0(t)&=\sum_{< i\alpha,j\alpha'>} t_1 e^{iA_{ij}}c_{i\alpha}^{\dagger}c_{j\alpha'} + M \sigma ^z_{\alpha \alpha'} c_{i \alpha}^{\dagger}c_{i\alpha'},
\label{eq:1a} \\
H(t)&=H_0(t)+U_{\text{dis}}
\label{eq:1b}
\end{align}
\end{subequations}
where $\alpha \in \{1,2 \}$ indicates sublattices A and B, $A_{ij}=\frac{e}{\hbar}{\bf A}(t)\cdot({\bf r}_i-{\bf r}_j)$ and $\vec{A}=A_0(\sin(\Omega t),\cos(\Omega t))$ is the vector potential for the incident circularly polarized light of frequency $\Omega$.
We consider nearest neighbour hopping with magnitude $t_1$. $\sigma ^z$ is the Pauli matrix, and $M$ is the staggered sublattice potential.
$H_0(t)$ represents the clean limit for the system and $H(t)$ is the full Hamiltonian with $U_{\rm dis}$ the disorder potential.
The disorder is chosen as an on-site chemical potential, and is diagonal in the real-space representation.
We choose the natural system of units $\hbar=e=c=1$ and set lattice spacing $a=1$.  The bandwidth of the time-independent part of $H_0(t)$ is $W$. As we explain below, the model of Eq. (\ref{eq:1}) can be directly implemented in the photonic lattice realization considered by Rechtsman et al. \cite{Rechtsman2013}.

The idea behind our construction of a Floquet topological Anderson phase is the following. A honeycomb lattice with a staggered potential , Eq. (\ref{eq:1}), has a gap $M$ at both Dirac cones.
A periodic drive alone also induces a gap, with masses of opposite sign at the two Dirac cones.
To second order, this gap is simply $\pm A_0^2v_F^2/\Omega$, for the $K$ and $K'$ points.
Thus, the drive induces effectively a Haldane model \cite{PhysRevLett.61.2015}, and yields an example
of a Floquet topological phase \cite{PhysRevB.84.235108,PhysRevB.79.081406}.
For weak and high-frequency ($\Omega\gg t_{1}$) drives, where perturbation theory is valid,
the drive and the staggering compete. Thus, the system is topological when $M<v_F^2 A_0^2/\Omega$, with a Chern number $|C_F|=1$, and trivial otherwise.
The key is the effect of disorder:  it diminishes a band gap induced by the drive, but even more strongly it suppresses the staggering.
Starting from the trivial phase, $M>v_F^2 A_0^2/\Omega$, an increase in disorder may reverse this balance, and induce a topological phase (for a static analog, see  Ref. \cite{PhysRevB.84.235108}).
In \cite{supp}, we provide a Born-approximation analysis of the disorder effects on the two gaps in
the static limit.

The explanation above, however, relies on weak, high frequency drive, which effectively produces a static perturbation. It does not capture the scenario in which the topological properties of the time-dependent system are a result of a resonance, connecting states of the original bulk band structure. In addition, we find that it is necessary to consider strong driving in order to observe the disorder induced topological phase. Below, we will establish the existence of the Floquet topological Anderson phase beyond the limit of a weak, high frequency drive. We will consider strong periodic drives, and will analyze two distinct frequency regimes: the high frequency regime ($\Omega >W$), and the low frequency ($\Omega<W$)  regime in which resonances occur within the band-structure. We will compare the two regimes and show that both of them exhibit a disorder-induced FTAI phase.

First, let us transform the problem defined in Eq. (\ref{eq:1}) into a time-independent Hamiltonian.
We define $H^F$ as follows:
\begin{equation}
H^F_{nm}=n\Omega\delta_{nm}+\int_0^{2\pi/\Omega}dt e^{i\Omega(n-m)t}H(t)
\label{eq:3}
\end{equation}
The 'Floquet' indices $n$ (and $m$) refer to replicas of the Hilbert space 
The eigenstates of  $H^F$ are the quasi-energies ($\epsilon$), which are
periodic in a quasi-energy "Brillouin" zone with period $\Omega$. We set the boundaries of this zone at $\pm\Omega/2$.
The off-diagonal terms (in Floquet indices) of  $H^F_{nm}$ emerge from the hopping term in Eq. (\ref{eq:1}), $\left(H_0\right)_{ij}=t_1\exp(iA_0 \cos(\Omega t+\phi_{ij}))$, where $(i,j)$ indicates hopping from site $i$ to $j$ and $\phi_{ij}=\pm\frac{\pi}{3}$ or $0$. Therefore,  $\left(H^F_{m,m+n}\right)_{ij}=t_1i^{n}J_{n}(A_0)\exp(i\phi_{ij})$, where $J_n(A_0)$ are the Bessel functions of the first kind.
Here, to efficiently use exact-diagonalization, we neglect $H^F_{m,m+n}=0$ for $n\geq 2$.
We also truncate $\left(H_F\right)_{nm}$, such that the Floquet indices obey $|n|,|m|\leq n_{max}$, with $n_{max}$ determined through convergence tests. The typical quasi-energy spectrum of our model is given in Figs. \ref{fig:1} (b) and (c), where we have defined a renormalized hopping, $\tilde{t}=t_1 J_0(A_0)$.

The quasi-energy band structure encodes the topological properties of time-periodic Hamiltonians.
While non-interacting equilibrium 2D Hamiltonians with broken time-reversal symmetry are classified by the Chern number, periodically-driven systems require a more general topological invariant - the winding number - which counts the number of edge states at a particular quasi-energy \cite{PhysRevX.3.031005}.   In disordered time-independent
systems, the disorder-averaged Chern number  is the Bott index, as defined by Hastings and Loring \cite{0295-5075-92-6-67004}. For our periodically-driven model, the disorder-averaged  winding number is calculated using the Bott indices obtained from the eigenvalues and eigenvectors of $H_F$, defined in Eq. (\ref{eq:3})
, and truncated to a finite number of replicas (for full details, see \cite{supp}). The Bott index at a particular quasi-energy, $C_b(\epsilon)$, for the truncated $H_F$, is the number of edge states at that quasi-energy \cite{PhysRevX.3.031005}. Also, the Chern number of a quasi-energy band is simply the difference in the Bott indices at the band edges.
\begin{figure}
\includegraphics[width=\linewidth]{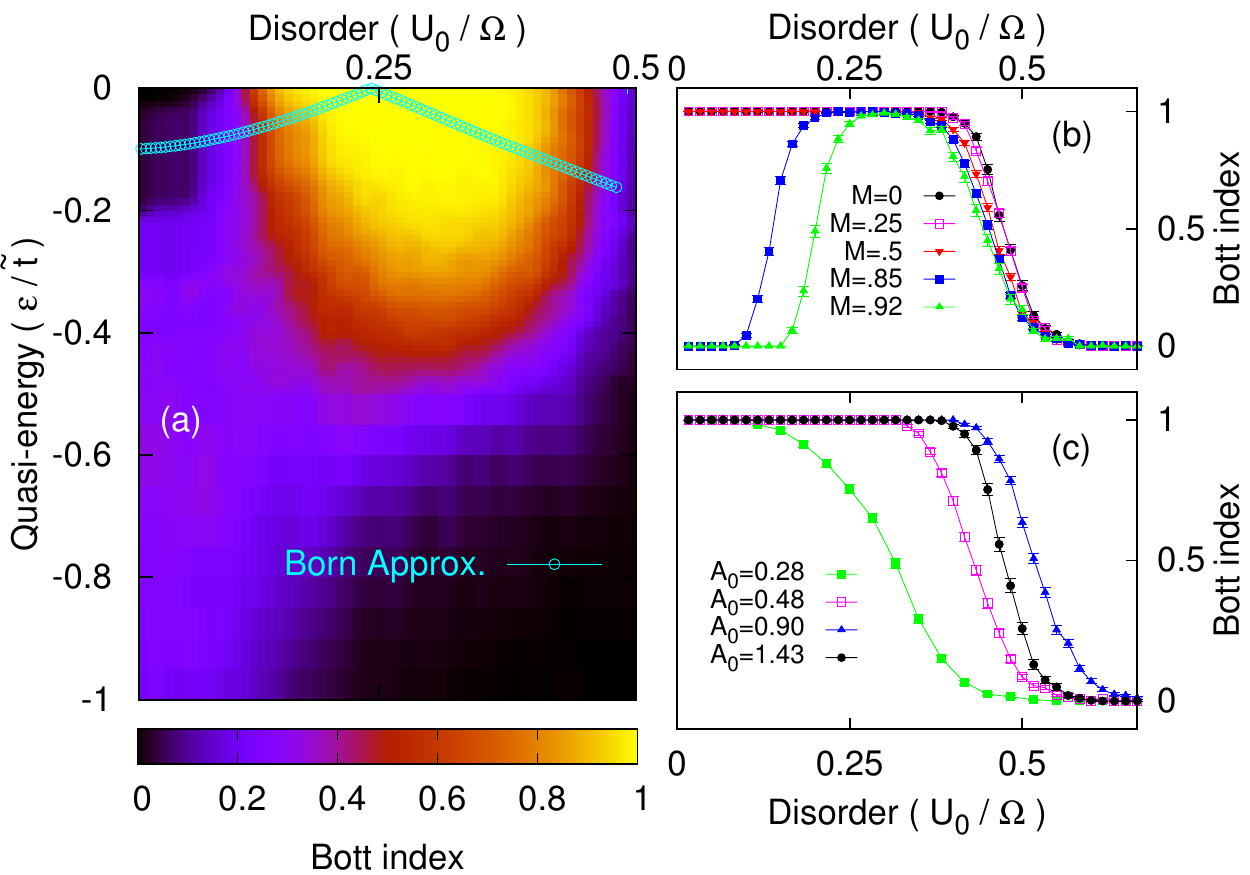}
\caption{(a) The Bott index,$C_b$, (in color), as a function of the quasi-energy and the disorder strength. Edge states are observed in the region where  $C_b(0)=1$. The quasi-energy gap,in the Born approximation, is shown(cyan) as a function of disorder.
The system parameters are $A_0=1.43$, $M/\tilde{t}=0.85$ and the size is $(L_x,L_y)=(30,30)$.
(b) $C_b$ as a function of disorder for different staggered masses $M/\tilde{t}=0$, $0.5$, $0.85$, $1$ at quasi-energy $\epsilon=0$ keeping $t_1$ and $A_0$ same as (a).  (c) $C_b$ as a function of disorder for different driving strengths, $A_0=0.28$, $0.48$, $0.90$ and $1.43$, keeping fixed $t_1$ and $M/t_1=0$. We have set $\Omega/t_1=12 J_0(1.43)$.
}
\label{fig:2}
\end{figure}

Let us first consider the case of $\Omega>W$ without resonances.
The clean system forms a trivial insulator, with its quasi-energy spectrum shown in Fig. \ref{fig:1} (c). The Bott index, $C_b$, as a function of disorder strength, $U_0$, and quasi-energy is shown in Fig. \ref{fig:2} (a).
At very weak disorder, the index, $C_b(\epsilon=0)=0$ in the quasi-energy gap, and it is not quantized at other quasi-energies, indicating a trivial phase.
A topological phase emerges as disorder increases, and is manifested by the Bott index becoming one, $C_b(0) \sim 1$.
This phase is induced by both disorder and drive, and, therefore, we identify it as a Floquet topological Anderson insulator (FTAI).
As expected, varying $M$ while keeping the drive strength fixed, shifts the position of the trivial-topological transition (see Fig. \ref{fig:2} (b)).
A qualitative description of this transition is provided by the disorder-averaged Born approximation \cite{supp}. Even though, this approximation (Fig. 2(a)) captures this basic physics of the transition, it overestimates the exact point of the transition. 

At  disorder strengths that are considerably larger than the transition point, the FTAI phase is destroyed and there is localization at all quasi-energies. This transition is insensitive to the staggered potential strength, as is evident from Fig. \ref{fig:2} (b); however, it depends on the drive strength (see Fig. \ref{fig:2} (c)). To observe the FTAI, the trivial-to-topological transition must occur well before the localization transition. Thus we consider the effects of strong driving (where $A_0 \sim 1$). As discussed in \cite{supp}, the finite-size dependence of the Bott index as a function of quasienergy in the topological phase is in agreement with the presence of an extended state in the bulk. The topological phase is protected against disorder if there is a `mobility  gap' in the spectrum, and some states are delocalized.

\begin{figure}
\includegraphics[width=\linewidth]{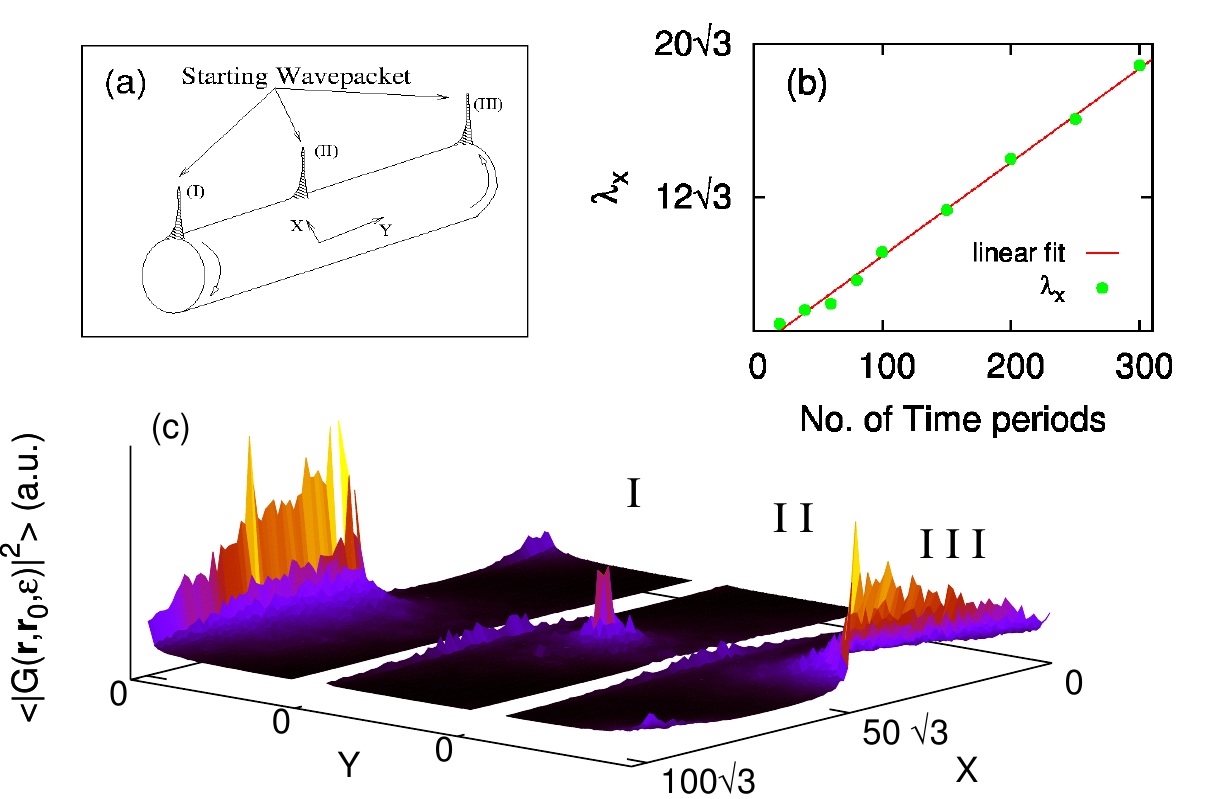}
\caption{(a) The cylindrical geometry for the time evolution of a starting $\delta$-function wavepacket. Cases (I), (II) and (III) have the starting positions, ${\bf r}_0\equiv(x_0,y_0)$ in the A sublattice at the left edge, bulk and right edge with $y_0/a_y=0$, $N_y/2-1$ and $N_y$  respectively. In all the cases, we fix $x_0/a_x=N_x/2$.
(b) The spread of $g_N({\bf r},{\bf r}_0,0)$ as a function of total time of evolution $T_f=NT$ along the $X$ direction, for ${\bf r}_0$ corresponding to case (I). $\lambda_x(N)$ grows linearly, with a velocity $v_{\rm edge}=(0.09\pm 0.001) a/T$.  (c) $g_N=\langle |G_N({\bf r},{\bf r}_0,\epsilon=0)|^2\rangle$
in real space as a function of ${\bf r}$, for the three cases, with $N=300$ and averaged over 400 realizations of disorder . Each sublattice has $N_x\times N_y=100\times30$ points. The system parameters are $A_0=1.43$, and $M/\tilde{t}=0.85$.
}
\label{fig:3}
\end{figure}

Next we numerically examine the existence of edge states as a diagnostic for topological phases.  The time-evolution operator for $H(t)$ is obtained in discrete time steps, $\delta t$ using a split-operator decompositon.
The honeycomb lattice [Fig. \ref{fig:1} (a)] is considered in a cylindrical geometry, with periodic boundary conditions along $X$ and open ones along $Y$ (see Fig. \ref{fig:3} (a)).
Initializing with a $\delta$-function wavepacket at ${\bf r}_0\equiv(x_0,y_0)$, the Green function, $G({\bf r},{\bf r}_0,t)$, is obtained from the time-evolution operator, $U(t,0)$.
An evolution for $N$ time periods ($T=2\pi/\Omega$) yields $G_N({\bf r},{\bf r}_0,NT)=\langle {\bf r}|U(t=NT, 0)|{\bf r}_0\rangle$.
The initial position, ${\bf r}_0$, is chosen to probe edge or bulk.
Compared to the analysis by exact-diagonalization of $H^F$, in this method we do not need approximations, and large system sizes are accessible.

The propagator, $G_N({\bf r},{\bf r}_0,NT)$ is the Floquet Green's function obtained from $H^F$ \cite{supp}.
So, the quasi-energy eigenvalues and eigenstates are analyzed by Fourier transforming the Green's function in time, $G_N({\bf r},{\bf r}_0,\epsilon)$.
With disorder, we calculate, $g_N({\bf r},{\bf r}_0,\epsilon)=\langle|G_N({\bf r},{\bf r}_0,\epsilon)|^2\rangle$,
where $\langle.\rangle$ indicates disorder averaging.
The extended or localized nature of the states at quasi-energy $\epsilon$ is given by the spread of $g_N$ defined as $\lambda_{x}(N)$, and $\lambda_{y}(N)$, along $X$ and $Y$ directions respectively.

The time-evolution is carried out for a system with $A_0=1.434$, $M/\tilde{t}=0.85$, $U_0/\tilde{t}=3.5$, and $\Omega/\tilde{t}=12$.
These parameters correspond to a FTAI and, thus we expect ballistic edge states at $\epsilon=0$.
The initial wavepackets are chosen in the A sublattice, on the two edges (cases (I) and (III)), and the bulk (II), as shown in Fig. \ref{fig:3} (a). After evolution for $N$ cycles,
 $g_N({\bf r},{\bf r}_0,0)$, for all three cases is shown in Fig. \ref{fig:3} (c). For cases (I) and (III), $g$, is extended along $X$ and localized in $Y$, indicating the presence of an edge state. The decay of $g_N$ along $X$ after some finite distance is due to finite time-evolution.
The chiral nature of the edge states are also revealed by the direction in which $g_N({\bf r},{\bf r}_0,\epsilon)$ evolves as a function of $N$.
Fig(\ref{fig:3})(b) shows that $\lambda_x(N)$ increases linearly with time of evolution, $N$, indicating that the edge states are ballistic and do not backscatter from impurities.
In contrast, bulk states are diffusive in nature until Anderson localization sets in. A finite amplitude is observed on the edge when starting with a bulk wavepacket because the bulk localization length is larger than the width of the system, indicating an overlap of the edge state wavefunction with the initial wavepacket.
We have shown the presence of protected edge states. This confirms the existence of the FTAI.
\begin{figure}
\includegraphics[width=\linewidth]{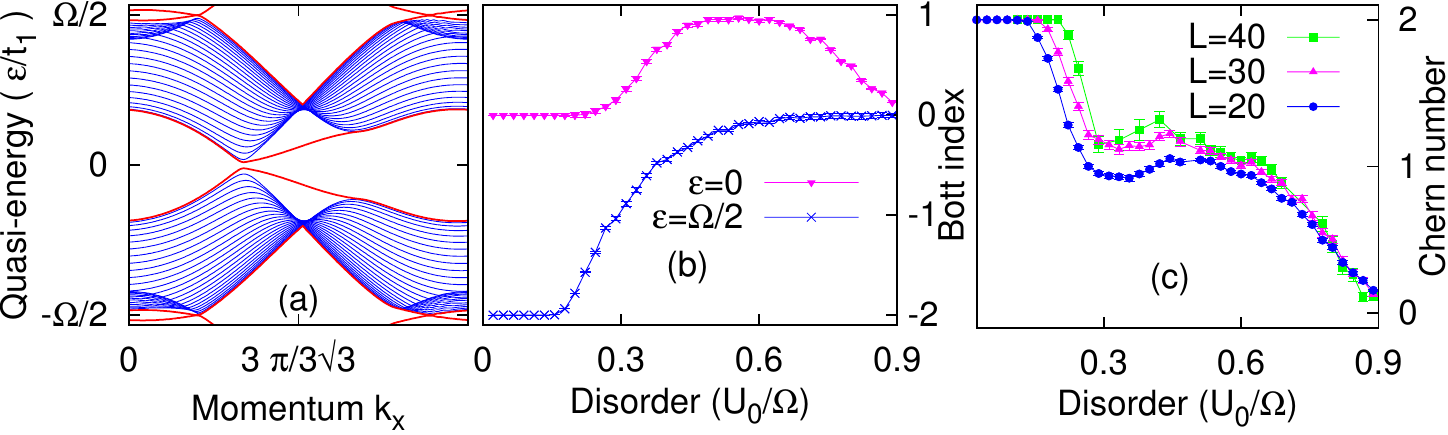}
\caption{(a) Band structure for the case of a single resonance. Edge-states (shown in red) are observed at the two bulk band gaps at
quasi-energies $\epsilon/t_1=0$ and $\Omega/2$. The gap at $\epsilon=0$ is made trivial by a staggered mass.
The system parameters are $A_0=0.75$, $M/t_1=0.3$, and $\Omega/t_1=9/2$.
(b) Disorder-averaged Bott index at a particular quasi-energy gap, $C_b(0)$ (magenta) and $C_b(\Omega/2)$ (blue).The Floquet Hamiltonian is truncated after 9 Floquet bands. System size is $(L_x,L_y)=(30,30)$. (c) Disorder-averaged Chern number, $C_F=C_b(0)-C_b(\Omega/2)$, of a single Floquet band between $\epsilon=-\Omega/2$ and $\epsilon=0$.
\label{fig:4}
}
\end{figure}

This novel phase persists even when there is a resonance within the band structure($\Omega<W$).
There, a transition occurs between an FTI phase and the disorder-induced FTAI phase.
Furthermore, the FTAI phase in this case cannot be understood using perturbative arguments since the resonance alters the topological nature of all the Floquet bands in the problem \cite{supp}.
Fig. \ref{fig:4} (a) shows the quasi-energy spectrum of the clean system.
The gap at the resonance, $\epsilon=\Omega/2$ is topological with $|C_b(\Omega/2)|=2$ and, thus, supports two edge states.
The gap at the Dirac points is trivial,   $|C_b(0)|=0$, since the staggered mass $M$ still dominates over the effect of the drive near $\epsilon=0$.
Fig. \ref{fig:4} (b) shows two transitions as disorder is increased.
A topological-to-trivial transition removes the edge states in the gap at the resonance ($\epsilon=\pm \Omega/2$).
Another transition induces topological edge states at $\epsilon=0$. From the finite sizes investigated, the topological to trivial transition at $\epsilon=\Omega/2$ happens initially and is unrelated to the transition at $\epsilon=0$.
Finally, disorder becomes strong enough to localize the entire band, as in the high-frequency case. The  Chern number of the band between these two quasi-energies, $C_F=C_b(0)-C_b(\Omega/2)$ changes from $|C_F|=2$ to  $|C_F|=1$, and then to $|C_F|=0$ (Fig. \ref{fig:4} (c)).
The intermediate regime, with $|C_F|=1$, is again identified as a FTAI - it is a topological state that requires both disorder and a periodic drive.
The fact that this phase exists even in a system which is non-perturbatively affected by the periodic drive indicates the universality and robustness of the FTAI.

This FTAI phase is directly amenable to experimental observation.  Recently a topological band-structure was experimentally demonstrated \cite{Rechtsman2013} in a structure composed of an array of coupled waveguides (a "photonic lattice").
  There, the diffraction of light is governed by the paraxial Schr\"odinger equation, wherein the spatial coordinate along the waveguide axis acts as a time coordinate. 
The guided modes of the waveguides are analogous to atomic orbitals, and  thus, the diffraction is governed by a tight-binding model. 
By fabricating the waveguides in a helical fashion, $z$-reversal symmetry is broken, resulting in a photonic Floquet topological insulator \cite{Lindner2011}, with topologically-protected edge states. 

The same system may give a realization of Eq.~(\ref{eq:1}) and the proposed FTAI phase. The gauge field, $A_0$, in the photonic system
is determined by the helix radius and period.  The sublattice potential, $M$, and on-site disorder, $U_0$, may be implemented by fabricating waveguides of different refractive indices, which is straightforwardly done in the laser-writing fabrication process \cite{Szameit}. Since each waveguide can be fabricated with a specified refractive index, the mass, $M$ and disorder strength, $U_0$ can be tuned entirely independently. In the supplementary section \cite{supp}, we fully discuss the relevant experimental parameters in the photonic lattice setup and demonstrate that the data we have presented here (shown in Figs. 2-4) are entirely amenable to experiment.  The topological transition may be probed by measuring transmission through the photonic lattice for samples of different disorder strengths.  For small disorder, the presence of a bulk band gap will give rise to zero transmission through the sample.  For disorder strengths above the transition, the presence of edge states in the band gap will allow 
transmission through 
the sample: a direct experimental observable.  Therefore, the FTAI phase may be implemented using an optical wavefunction in a photonic crystal structure, as opposed to an electronic wavefunction in a condensed matter system.

In summary, we have established the existence of a disorder-induced Floquet Topological Insulator phase. Starting from a clean system that is trivial
even in the presence of time-periodic driving, disorder renormalizes the parameters of the Hamiltonian to make they system topological.
Experimentally the parameters are in a range that can be achieved in a photonic lattice, and this could be a first experimental
realization of the Topological Anderson Insulators.

We thank Kun W. Kim, Shu-Ping Lee, Victor Chua and S. M. Bhattacharjee for illuminating discussions. This work was funded by the Packard Foundation, and by the Institute for Quantum Information and Matter, an NSF Physics Frontiers Center with support of the Gordon and Betty Moore Foundation. NL acknowledges financial support by the Bi-National Science Foundation, and by ICore: the Israeli Excellence Center "Circle of Light".

 \bibliographystyle{apsrev}

\end{document}


\begin{center}
 \large{\bf SUPPLEMENTARY MATERIALS}
 \end{center}

 \section{Floquet-Bloch Theory: Definitions}
Let us start with the Hamiltonian $H(t)$ that is periodic in time,
\begin{eqnarray}
 H({\bf k},t)&=&H_0({\bf k})+V(t), \\ \label{seq:1}
H({\bf k},t)&=&H({\bf k},t+T), \ \ {\rm with} \ \ T=2\pi/\Omega, \nonumber
\end{eqnarray}
as the time-period,  $\Omega$ being the frequency. Here, $H_0$ contains the time-independent terms of the Hamiltonian. The states are given by the solution to the full time-dependent Schr{\"o}dinger equation,
\begin{equation}
i\hbar \frac{\partial}{\partial t}\psi({\bf k},t)=H(t) \psi({\bf k},t)\label{seq:1.5}
\end{equation}
The Floquet-Bloch theorem states that, the time-evolution operator can be written as
\begin{equation}
U(t,0)=\exp\left(-iH^Ft\right)W(t),\ \ {\rm with} \ \ W(t+T)=W(t), \label{seq:2}
\end{equation}
and $H^F$ is a time-independent Hermitian operator. The form of Eq. (\ref{seq:2}) allows us to identify $H^F$ as an effective time-independent Floquet Hamiltonian.

In order to define $H^F$, for the case at hand, the Fourier decomposition of the solution to Eq. (\ref{seq:1.5}) is used, 
\begin{eqnarray}
\psi({\bf k},t)&=&\sum_n \psi_n({\bf k}) e^{i n \Omega t},\label{seq:3}\\
&=&\sum_n \langle n|\psi^F\rangle \langle t|n\rangle, \ \ {\rm with}\ \ \langle t|n\rangle =e^{i n \Omega t}. \label{seq:4}
\end{eqnarray}
In Eq. (\ref{seq:4}), we have introduced an additional register particle, $\{|n\rangle\}$,  where $n \in \mathds{Z}$. The Floquet Hamiltonian, $H^F$, is defined in a way such that $|\psi^F\rangle$ are eigenstates. Necessarily, it is defined in an extended Hilbert space $\mathscr{H} \otimes \{|n\rangle\} $, where $\mathscr{H}$ is the original Hilbert space of the Hamiltonian (see Eq. (\ref{seq:1})). The time-dependent Schr{\"o}dinger equation is rewritten in an effective time-independent form,
\begin{equation}
H^F|\psi^F\rangle =\epsilon |\psi^F \rangle,\label{seq:5}
\end{equation}
where $H^F$ is infinite dimensional.  The eigenvalues ($\epsilon$) are referred to as quasi-energies, and the eigenfunctions of the Floquet Hamiltonian, defined in Eq. (\ref{seq:5}), are the quasi-energy states.
 The spectrum of $H^F$ is unbounded; however, we note that in Eq. (\ref{seq:2}), the eigenvalues ($\epsilon$) of $H^F$ describe the non-periodic evolution of these states as a function of time. Therefore, they are unique modulo $\Omega$, $\epsilon \equiv \epsilon +m\Omega$. 
The explicit form of the Floquet Hamiltonian for $H(t)$ defined in Eq. (\ref{seq:1}) is,
 \begin{eqnarray}
 & (H^F({\bf k}))_{mn}&\equiv\langle m|H^F({\bf k})|n\rangle, \nonumber\\
& &=(H_0({\bf k})+n\Omega) \delta_{mn}+\tilde{V}_{mn}, \label{seq:6} \\
{\rm where,} & & \nonumber \\
&\tilde{V}_{mn}&= \frac{1}{T} \int_0^T dt V(t)e^{i(m-n)\Omega t }
 \end{eqnarray}
The integers $m$ and $n$ indexes a particular Floquet block in the matrix $H^F$. In this representation, the time-independent terms, like $H_0$, are diagonal, but the time-dependent potential, $V(t)$, acts as a hopping amplitude between various Floquet blocks. These Floquet blocks are like replicas of the original Hamiltonian shifted in quasi-energy by $\Omega$, and the indices will also be referred as the replica index.
The quasi-energies can be computed by truncating the matrix after a certain number of Floquet blocks and diagonalizing it.

The Floquet Green function is defined as 
\begin{equation}
 G^F=\frac{1}{(E\mathds{1}-H^F)}. \label{seq:7}
\end{equation}
All elements of $G_F$ can be rewritten in a closed analytical formula \cite{Martinez} for the special case where the only non-zero components of $\tilde{V}_{mn}$ are $V_-=\tilde{V}_{m+1,m}$ and $V_+=\tilde{V}_{m,m+1}$ with $m \in \mathds{Z}$ . We mostly restrict ourselves to the $(0,0)$ Floquet block,
\begin{eqnarray}
& &(G_F)_{00}=\frac{1}{E\mathds{1}-H_0-V_{\rm eff}^+-V_{\rm eff}^-} \label{seq:8}, \\
& &V_{\rm eff}^\pm=V_+\frac{1}{E\pm \Omega-H_0-V_+\frac{1}{E\pm 2 \Omega -H_0-V_+\frac{1}{\vdots} V_-}V_-}V_-.\nonumber
\end{eqnarray}
The Green function can be obtained perturbatively to any order in $V$ by truncating the continued fraction at that order. 
\section{Floquet Topological Insulators: Haldane model and higher Chern insulators}
 The topological behavior in the non-equilibrium situation is obtained by choosing a drive of appropriate frequency. 
We show the non-trivial topology of the quasi-energy band-structure for the graphene based model in the presence of circularly polarized light. 

The tight-binding model on a hexagonal lattice with nearest neighbor hopping and without radiation, in the low energy and linearized momentum regime reduces to
\begin{equation}
H_0=v_F(k_x\sigma_z\tau_z+k_y\sigma_y)+M\sigma_z \label{seq:10}
\end{equation}
where $\sigma_x$ and $\tau_z$ refer to sub-lattice isospin and valley degree of freedom respectively, $v_F$ is the Fermi velocity at the Dirac points and $M$ is the sub-lattice mass term. In the presence of circularly polarized light, using Pierels substitution, we have
\begin{eqnarray}
H(t) &=& v_F((k_x-A_x)\sigma_x\tau_z+(k_y-A_y)\sigma_y) +M\sigma_z \label{seq:11} \\
{\bf A}(t)&=&A_0(sin(\Omega t),cos(\Omega t)) \nonumber
\end{eqnarray}
where ${\bf A}$ is the vector potential for incident radiation. Consider a general form of the external drive defined in Eq. (\ref{seq:1}),
\begin{equation}
V(t)=V_+e^{i\Omega t}+V_-e^{-i\Omega t}, \label{seq:9}
\end{equation}
where $V_\pm$ are time-independent operators. Therefore, in the model considered, we have,
\begin{eqnarray}
V_+&=&A_0\left(\frac{i}{2}\sigma_x\tau_z-\frac{1}{2}\sigma_y \right),  \label{seq:11.5}\\
V_-&=&V_+^\dag. \label{seq:11.75}
\end{eqnarray}
Note that the analysis discussed here (see Eq. (\ref{seq:10}) to Eq. (\ref{seq:11.75})) is valid only in the perturbative low energy regime with $|{\bf A}|\ll 1$.

This model breaks time-reversal symmetry, and is classified by the Chern number. We explore two cases, (a) zero resonances, and (b) a single resonance due to radiation and their effects on the topology. 
\subsection{(A) No resonances}
 This case corresponds to irradiating the system with off-resonant light. The incident frequency of the drive, $\Omega \gg W$, where $W$ is the bandwidth of the time-independent band-structure. The correction to the energies of the non-equilibrium states are obtained by inspecting the poles of the Floquet Green function. In this case, to lowest order in the radiation potential, the off-diagonal terms in $G^F$ can be ignored. The diagonal element, $G^F_{00}$, to $O(V^2)$ is,
\begin{equation}
 G^F_{00}=\left(E\mathds{1}-H_0-\frac{[V_+,V_-]}{\Omega}\right)^{-1}=(E\mathds{1}-H_{\rm eff})^{-1}
 \label{seq:12}
\end{equation}
where we have a new effective Hamiltonian, $H_{\rm eff}$. Using equations (\ref{seq:11.5}), (\ref{seq:11.75}) and (\ref{seq:12}), we note that $H_{\rm eff}$ is equivalent to the Haldane model for anomalous quantum Hall effect with a topological mass $\Delta_0=\frac{v_F^2 A_0^2}{\Omega}$,
\begin{eqnarray}
& & H_{\rm eff}=v_F(k_x \sigma_x \tau_z +k_y \sigma_y)+M\sigma_z+\Delta_0\sigma_z\tau_z \label{seq:13} \\
&=&\left(\begin{array}{cccc}
\Delta_+ & k_x-ik_y & 0 & 0\\
k_x+ik_y& -\Delta_+ & 0&0 \\
0 & 0 & \Delta_-& -k_x-ik_y\\
0 & 0 & -k_x+ik_y& -\Delta_-
\end{array}\right),
\end{eqnarray}
 where $\tau$ denotes the valley space, and $\Delta_{\pm}=M\pm\Delta_0$. The mass gap opens at the Dirac points of the band-structure near $\epsilon=0$.
The bands will be topological or trivial when $M<\Delta_0 $ and $M>\Delta_0$ respectively. Specifically, for $M<\Delta_0$, the Chern number, $C_n=1$, when measured at quasi-energies in the gap, $-(\Delta_0-M)<\epsilon<\Delta_0-M$, and is zero at all other quasi-energies.
\subsection{(B) Single resonance}
This scenario corresponds to the driving frequency in the regime $W/2<\Omega<W$. 
The quasi-energy band-structure has two gaps at (i)$\epsilon=0$, and (ii) $\epsilon=\pm \Omega/2$, where the topologically non-trivial features may be measured. 
The gap at $\epsilon=0$ is the same as that discussed in case (A) and is equal to $\Delta_0$. We incorporate the effect of off-resonant processes on the quasi-energy band-structure by making the replacement $H_0 \rightarrow H_{\rm eff}$ in the Floquet Hamiltonian defined in Eq. (\ref{seq:6}).
For quasi-energies close to resonance, $\epsilon\sim \Omega/2$, adjacent diagonal Floquet blocks, $H_{\rm eff}$, and $H_{\rm eff}-\Omega$, are nearly degenerate. 
Therefore, to lowest order, $H_F$ must be diagonalized in this subspace  of two adjacent Floquet blocks, to obtain the correct quasi-energies. 
The effective two band Hamiltonian is given by,
\begin{equation}
(H^F)_{\rm eff}= P_\Omega\left(\begin{array}{cc}
H_{\rm eff}-\Omega & V_+ \\
V_- & H_{\rm eff} \end{array} \right)P_\Omega,
\end{equation}
where $P_\Omega$ is the projector onto the bands with quasi-energies in the range $0<\epsilon<\Omega$.
This is exactly the same as degenerate first order perturbation theory, and therefore, the gap exactly at resonance, $\epsilon=\Omega/2$, is proportional to $|V_\pm|$.

The quasi-energies of $H^F$ are periodic in $\Omega$. 
To properly define a the Chern number for a band ($C_n$), we must specify its upper and lower bound in quasi-energies. 
An alternative is to measure the Chern number ($C_n^{\rm trunc}$) of all bands below a particular quasi-energy, for a truncated $H^F$. 
It has been shown \cite{PhysRevX.3.031005} that this number corresponds to the number of edge states that will be observed at that particular quasi-energy irrespective of chirality.
 For the case of single resonance, the Chern number of the truncated $H^F$ for $M<\Delta_0$ is
\begin{equation}
C_n^{\rm trun}=\begin{cases}
1 &\mbox{if } \epsilon=0 \\
2 & \mbox{if } \epsilon=\pm\Omega/2,
\end{cases}
\end{equation}
and for $M>\Delta_0$
\begin{equation}
C_n^{\rm trun}=\begin{cases}
0 &\mbox{if } \epsilon=0 \\
2 & \mbox{if } \epsilon=\pm\Omega/2,
\end{cases}
\end{equation}
The Chern number of the  bands are $C_n=\pm3$ when $M<\Delta_0$, and $C_n=\pm2$ when $M>\Delta_0$.

\section{Born Approximation: Details}
 The transition from a trivial state to a topological state is due to renormalization of parameters of the Hamiltonian due to disorder.
 In the lowest order Born approximation, the correction to the density of states are obtained from exact analytical expressions for the self energy. This provides an accurate description for the density of states as a function of disorder at dilute disorder. The disorder averaged Floquet Green function is given by,
\begin{equation}
\langle G^F(i\omega_n,{\bf k}) \rangle= \frac{1}{i\omega_n-H^F({\bf k})-\Sigma(E)},
\end{equation}
and,
\begin{equation}
 \Sigma(i\omega_n,{\bf k})=\int_{{\rm FBZ}} d{\bf k'}\ \ \langle U_{\rm dis}({\bf k},{\bf k'})G^F(i\omega_n,{\bf k'})U_{\rm dis}({\bf k'},{\bf k})\rangle,
 \label{seq:23}
\end{equation}
where, $\langle \dots \rangle$ denotes disorder averaging, $U_{\rm dis}({\bf k},{\bf k'})$ is the disorder potential in Fourier space, and $\Sigma$ is the self energy. We are interested at the physics of the topological transition near $\epsilon=0$ as a function of disorder. For the case of zero resonances, this is correctly modeled by the effective Hamiltonian, $H_{\rm{eff}}$, defined in Eq. (\ref{seq:13}). Therefore, instead of using the Floquet Green function, $G^F$, we use the effective Green function given by,
\begin{equation}
G_0^{\rm eff}(i\omega_n, {\bf k})=\frac{1}{i\omega_n-H_{\rm eff}({\bf k})}.
\end{equation}

The disorder potential, $U_{\rm dis}$, is modeled as $\delta$-correlated point scatterers. The short range of scattering implies that both inter- and intra-valley processes must be taken into account. It is assumed that, in the linearized regime, the disorder matrix in real space is \cite{JPSJ.67.2421},

\begin{equation}
U_{\rm dis}(\vec{r})=\sum_{i}\left(\begin{array}{cccc}
U_{i}^{A} & 0 & U_{i}^{A}e^{i\phi_{i}^{A}} & 0\\
0 & U_{i}^{B} & 0 & U_{i}^{B}e^{i\phi_{i}^{B}}\\
U_{i}^{A}e^{-i\phi_{i}^{A}} & 0 & U_{i}^{A} & 0\\
0 & U_{i}^{B}e^{-i\phi_{i}^{B}} & 0 & U_{i}^{B}
\end{array}\right)
\label{seq:20}
\end{equation}
where, 
\begin{eqnarray}
U_{i}^{A,B}&=&u_{i}^{A,B}\delta({\bf r}-{\bf r}_{i}^{A,B}), \\
\phi_{i}^{A,B}&=&({\bf K}'-{\bf K})\cdot{\bf r}_{i}^{A,B}. 
\end{eqnarray}
$A$ and $B$ refer to the different sub-lattices, ${\bf K}$ and ${\bf K'}$ are the two valleys, and $i$ is summed over the unit cells. 
The disorder potentials $u_i^{A,B}$ are taken from an uniform distribution in the range $[-U_0/2,U_0/2]$ and are $\delta$-correlated. Therefore,
\begin{eqnarray}
\langle u_{i}^{A}\rangle&=&\langle u_{i}^{B}\rangle=0, \\
\langle u_{i}^{\nu}u_{j}^{\nu'}\rangle&=&\frac{U_0^2}{12}\delta_{ij}\delta_{\nu\nu'}, \ \ \nu,\nu'\equiv A,B,
\end{eqnarray}
 where we have used that the variance of the uniform distribution is $U_0^2/12$. The diagonal and off-diagonal terms in Eq. (\ref{seq:20}) account for intra- and inter-valley scattering respectively and are assumed to have the same magnitude.

 The self energy can be calculated by rewriting $U_{\rm dis}$ (see Eq. (\ref{seq:20})) in Fourier space and using Eq. (\ref{seq:23}) . In the limit of $|{\bf k'}|\ll|{\bf K}-{\bf K'}|$, it can be assumed that, the fast oscillating exponents in the off-diagonal terms in the self energy Eq. (\ref{seq:23}) averages to zero \cite{JPSJ.67.2421}, i.e., 
 \begin{equation}
 \langle \sum_i e^{i ({\bf k'}.r_i^\nu \pm \phi_i^\nu)}\rangle=0.
 \end{equation}
 Therefore, the self energy is diagonal in valley space and independent of momentum ${\bf k}$.  Consequently, after integrating out the momentum, ${\bf k'}$, in the first Brillouin zone, the four main contribution to the self energy are,
\begin{equation}
\Sigma =\Sigma_I \mathds{I}+\Sigma_M\sigma_z+ \Sigma_\Delta \sigma_z\tau_z+\Sigma_0 \tau_z,
\end{equation}
with,
\begin{eqnarray}
&\Sigma_I& =-nu^{2}\frac{i\omega_{n}}{4\pi v_{F}^{2}}\log\left(\frac{v_{F}^{4}D^{4}}{f_+f_-}\right),\label{eq:8}\\
&\Sigma_M& = -\frac{nu^{2}}{4\pi v_{F}^{2}}\left[M\log\left(\frac{v_{F}^{4}D^{4}}{f_+f_-}\right)+\Delta \log \left(\frac{f_-}{f_+}\right)\right],\label{eq:9} \\
&\Sigma_\Delta& = \Sigma_0=0, \label{eq:10} 
\end{eqnarray}
where $f_\pm=\omega_{n}^{2}+(M\pm\Delta_0)^{2}$. Therefore, the parameters in $H_{\rm eff}(t)$ get renormalized as 
\begin{eqnarray}
i\tilde{\omega}_{n} &=&  i\omega_{n}-\Sigma_{0}^{I}, \\
\tilde{M}  &=&  M+\Sigma_{0}^{z}, \\
{\rm and},\ \ \tilde{\Delta_0}  &=&  \Delta_0. 
\end{eqnarray}
\begin{figure}
\includegraphics[width=\linewidth]{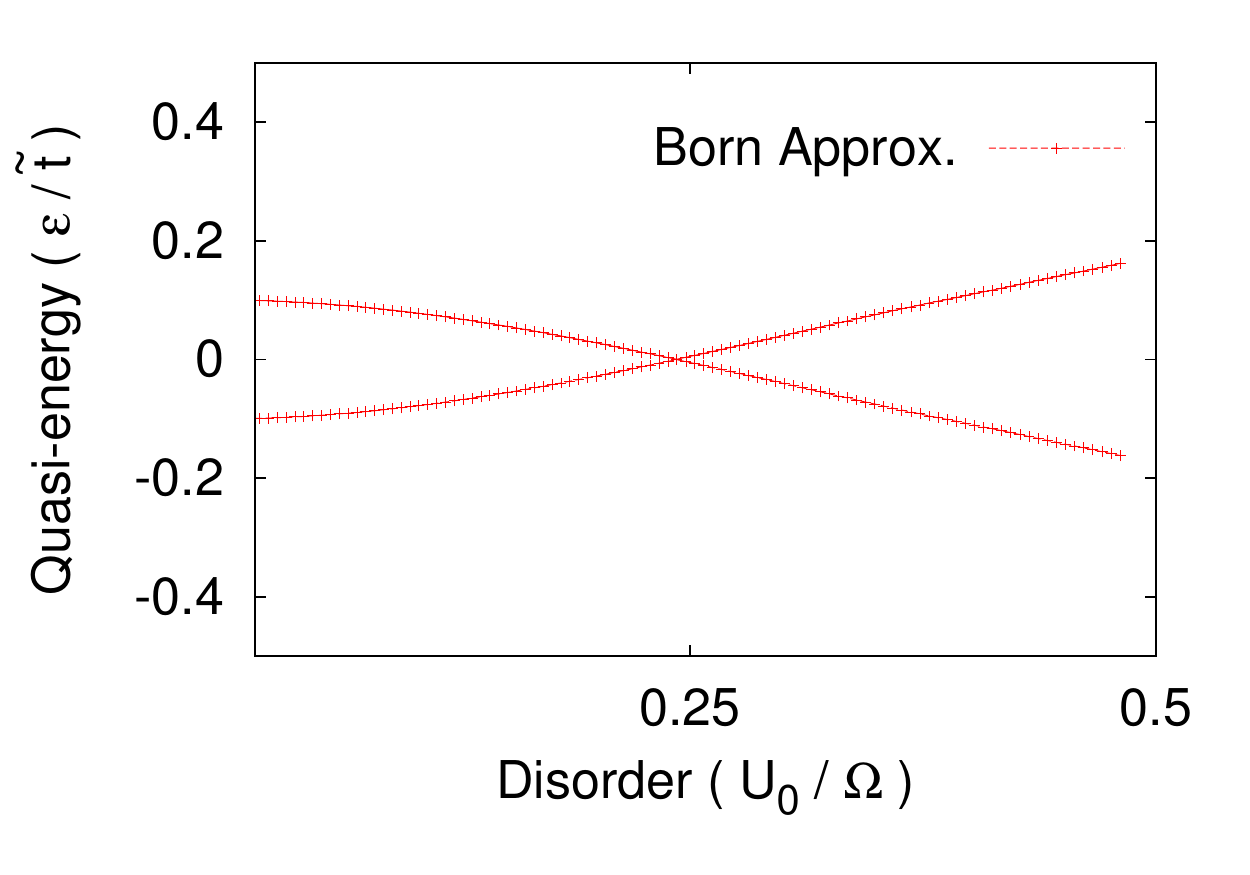}
\caption{The expected quasi-energy gap as a function of disorder given by the Born approximation. This is obtained by plotting the solution to the $\tilde{\omega}=\tilde{M}-\tilde{\Delta_0}$ as a function of disorder. 
The parameters for the system are $A_0=1.434$, $\Delta_0=0.75$ and $M=0.85$.}
\label{suppfig:1}
\end{figure}
The renormalized mass, $\tilde{M}$, reduces with increasing disorder. The renormalized quasi-energy is obtained by analytical 
 continuation of $i\omega_n \rightarrow \omega$ and the band gap as a function  of disorder is the solution to the equation $\tilde{\omega}=\tilde{M}-\tilde{\Delta_0}$. This is shown in Fig. (\ref{suppfig:1}) with parameters $v_f=3/2$, $\Delta=0.75$
  $M=0.85$ and $D=4\pi/3$. These parameters correspond to the case (I) of zero resonances. The topological phase transition occurs at the point where the band gap vanishes, which 
  happens when $\tilde{M}=\tilde{\Delta}$. For stronger disorder, the gap reopens in the topological phase and a non-vanishing Chern number must therefore be measured at the quasi-
  energies in the gap.
\begin{figure}[htbt]
\includegraphics[width=\linewidth]{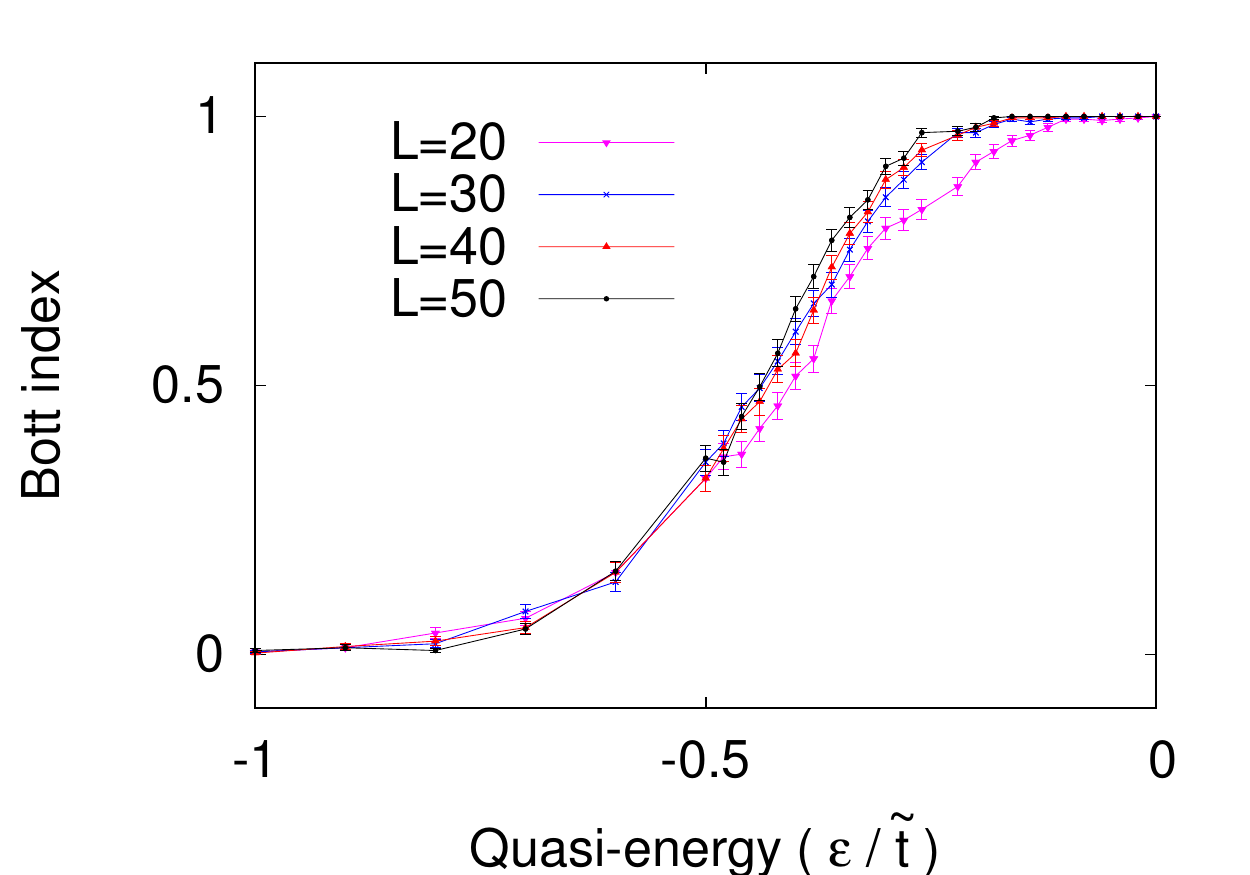}
\caption{Finite size effect of the disorder averaged Bott index at disorder strength, $U_0=3.5$, for the zero resonance case,as a function of quasi-energy for sizes 
$L_x=L_y=20$, $30$, $40$ and $50$. The index has been averaged over 400 disorder realizations. }
\label{suppfig:2}
\end{figure}

  \section{Bott Index for Floquet Hamiltonian.}
  We outline the method to obtain the Chern number of bands for disordered periodically driven Hamiltonians. The Bott index as a measure to obtain the Chern number, was defined by Hastings and Loring \cite{0295-5075-92-6-67004} for time-independent Hamiltonian. We generalize this formula for periodically driven systems by using the eigenstates of the truncated Floquet Hamiltonian, $H^F$. This index will measure the number of edge states at a given quasi-energy\cite{PhysRevX.3.031005}. Consider a Hamiltonian, $H(t)$ defined on a lattice with periodic boundary conditions (torus geometry). Given two diagonal matrices $X_{ij}=x\delta_{ij}$ and $Y=y\delta_{ij}$ for the $x$ and $y$ coordinates of the lattice sites, let us define two unitary matrices,
  \begin{eqnarray}
   U_X&=& \exp(i2\pi X/L_x),\\
   U_Y &=& \exp(i2\pi Y/L_y),
  \end{eqnarray}
where $L_{x,y}$ are the dimensions of the system. In the extended Floquet Hilbert space, the analogous definition for the unitary matrices are,
  \begin{eqnarray}
   \left(U^F_X\right)_{mn}&=& U_X \delta_{mn},\\
   \left(U^F_Y\right)_{mn} &=& U_Y \delta_{mn},
  \end{eqnarray}
  where $(m,n)$ refer to a particular Floquet block. For a band of quasi-energies, $\epsilon_l<\epsilon<\epsilon_h$, the Bott index is an integer, and  it is well defined as long as the lower and upper bounds, $\epsilon_{l,h}$, are in a mobility gap of the quasi-energy band-structure. The topological invariant is calculated using the unitary matrices projected onto a band. Let $P$ be the projector onto the chosen band of quasi-energies. In our system, we will be calculating the Bott index of all states with quasi-energies, $\epsilon <0$, in the truncated Floquet Hamiltonian, $H^F$. The projected unitary matrices are defined as,
  \begin{equation}
      \tilde{U}^F_{X,Y}= PU^F_{X,Y}P.
  \end{equation}
For a given disorder configuration, the Bott index of the band is given as,
\begin{equation}
 C_b=\frac{1}{2\pi}{\rm Im}\left[{\rm Tr}\left(\log\left(\tilde{U}^F_{Y}\tilde{U}^F_{X}\tilde{U}^{F\dag}_{Y} \tilde{U}^{F\dag}_{X} \right)\right)\right].
\end{equation}
The Bott index is a measure of commutativity of these projected unitary matrices, and it can be shown to be equivalent to the Kubo formula for the Hall conductivity\cite{0295-5075-92-6-67004}. For a given disorder strength, the Bott index must be averaged over a large number of configurations.

\section{Finite Size effect}
We investigate the finite size effect on the non-quantized region of the Bott index when in the Floquet-Anderson topological insulator (FATI) phase. 

In Fig. \ref{suppfig:2}, we have plotted the disorder averaged Bott index as a function of quasi-energy for different system sizes. It is clear that with increasing system size, the non-quantized region of the Bott index becomes sharper. This is in agreement with the expectation of a sharp extended state in the bulk quasi-energy band, analogous to a quantum hall state. Therefore, we expect the localization transition to be in the quantum Hall universality class. The current accessible system size is not sufficient to obtain the critical exponents of this transition.

\section{ Obtaining experimental parameters associated with the FTAI}

The equation describing the paraxial diffraction of light through an array of waveguides is a Schr\"odinger equation \cite{Rechtsman2013,Kawano-Kitoh}:
\begin{equation}
i\partial_z \psi = -\frac{1}{2k}\nabla_\perp^2\psi - \frac{k\Delta n (x,y,z)}{n_0} \psi,\label{Eq1}
\end{equation}
where $z$ is the distance of propagation along the waveguide axis; $k$ is the ambient wavenumber in the medium; $\nabla^2_\perp$ is the Laplacian in the transverse (x,y) plane; $n_0$ is the refractive index of the ambient medium, and $\Delta n(x,y,z)$ is the refractive index variation as a function of position that describes the waveguides.  Each waveguide in the lattice is best described by a hypergaussian function, where the refractive index pattern (for a single straight waveguide) can be written as $\Delta n^{(1)}(x,y,z) =\Delta n_1 \exp\left( -[(x/\sigma_x)^2 + (y/\sigma_y)^2]^3 \right)$.  The fabrication procedure referred to in the main text [26] allows for a range of values for $\Delta n_1$, depending on the speed of laser writing - taking values between $0.5\times 10^{-3}$ and $1.1\times 10^{-3}$.  The writing procedure leads to waveguides that are elliptical in shape (because the size of the focus of the writing beam in the transverse direction can be made much smaller than that in the longitudinal 
direction), with $\sigma_x \sim 2\mu m$ and $\sigma_y \sim 5.5\mu m$.  \\

In order to calculate coupling parameters between waveguides, a plane-wave expansion procedure \cite{Kawano-Kitoh,Johnson} is used in order to exactly numerically diagonalize Eq. (\ref{Eq1}).  This procedure is carried out for both a single waveguide and for two waveguides spaced at a distance $d$ from one another.  The former allows the calculation of the on-site energy (or ``propagation constant'' in optics terminology), which is associated with the chosen refractive index of the waveguide: this is just the eigenvalue associated with the bound mode of a single-mode waveguide.  The latter allows for the calculation of the evanescent coupling/hopping constant between two neighboring waveguides: this is just half the splitting between the eigenvalues of the two bound modes associated with the coupled waveguides.  \\

In order to fully examine the realizability of the experimental setup, we calculate - using standard numerical techniques \cite{Kawano-Kitoh} - that the hopping parameter can be tuned over an extremely large range ($0.083 cm^{-1}$ through $2.7 cm^{-1}$) because the nature of the coupling between adjacent waveguides is evanescent (these values correspond to the waveguides discussed in Ref. \cite{Rechtsman2013}, at lattice spacings $30\mu m$ and $12 \mu m$, respectively).  Furthermore, the on-site energies may be varied significantly by varying the refractive index difference of the waveguides relative to the background (which can be realistically varied in the range of $5.0 \times 10^{-4}$ through $1.1 \times 10^{-3}$ - as discussed above).  Assuming a waveguide helix pitch of $1cm$ as in Ref. \cite{Rechtsman2013}, the parameter $U_0/\Omega$ may vary over a range of $\sim \pm 1.8$ - this incorporates the full range of parameters discussed in the main section of this paper.  For a typical hopping parameter of 
$t_1=1.5cm^{-1}$, the dimensionless parameter $U_0/t$ (the degree of on-site disorder in units of the hopping) can take on values anywhere from $U_0/t = 0$ through $1.6$ - again, this includes the range discussed in the present paper.   

It is important to demonstrate that the strengths of the gauge field, $A_0$, as used here are directly realizable under experimental conditions.  As shown in Ref. \cite{Rechtsman2013}, the dimensionless expression for the strength of the gauge field is $A_0=kR\Omega a$, where $k$ is the wavenumber in the ambient medium ($k=2\pi n_0/\lambda$, where $n_0=1.45$ is the refractive index of fused silica, and $\lambda=0.633\mu m$ is the wavelength of laser light used); $R$ is the radius of the waveguide helices; $\Omega=2\pi/1cm$ is the spatial frequency of the helices, and $a=15\mu m$ is the lattice spacing between nearest neighbor waveguides.  In the experimental work \cite{Rechtsman2013}, the helix radius was tuned (with a constant helix pitch of $1cm$) from $0\mu m$ through $16\mu m$.  This corresponds to a gauge field strength (in the dimensionless units of the present paper) of $A_0=0$ through $A_0=2.17$.  The experimental work showed that this range was fully accessible experimentally.  In the present paper, 
we perform calculations at a number of different gauge field strengths, including 0.28, 0.48, 0.9 and 1.43 (for the calculations shown in Fig. 2); 1.43 (for the calculations in Fig. 3); and 0.75 (for the calculations in Fig. 4).  All of these are clearly experimentally accessible.   \\

Taken together, we have shown here that the parameters proposed in this work are directly amenable to experimental realization.\\ 

 \bibliographystyle{apsrev}